\documentclass[sigconf, 10pt]{acmart}

\usepackage{import}
\usepackage[caption=false,font=footnotesize]{subfig}
\usepackage{hyphenat}
\usepackage{xcolor}
\usepackage{amsmath}
\newcommand{\cmar}[1]{\textcolor{blue}{}}

\newcommand{\packetin}{\texttt{OFPT\_PACKET\_IN} }
\newcommand{\packetout}{\texttt{OFPT\-\_\-PACKET\_\-OUT} }
\newcommand{\echorequest}{\texttt{OFPT\_ECHO\_REQUEST} }

\newcommand{\featuresrequest}{\texttt{OFPT\_FEATURES\_REQUEST} }

\newcommand{\portstats}{\texttt{OFPC\_PORT\_STATS} }
\newcommand{\tcpnodelay}{\texttt{TCP\_NODELAY} }

\renewcommand\footnotetextcopyrightpermission[1]{} 
\begin{document}






%

\title{What it costs! Measuring the impact of multi-tenancy in virtual SDN networks}

\title{Identifying and Quantifying Interference in Virtual SDNs}

\title{Sources of Interference in Virtual SDNs}

\title{Logically Isolated? Actually Interfering!\\ Identifying and Measuring the Impact of Multi-Tenancy in Virtual SDNs}

\title{Measuring the Impact of Multi-Tenancy in Virtual SDNs}

\title{{\Large Logically Isolated? Actually Interfering!}\\ Measuring 
the Impact of Multi-Tenancy in Virtualized SDNs}

\title{{\Large Logically Isolated, Actually Unpredictable?}\\ Measuring
Hypervisor Performance\\  
in Multi-Tenant SDNs}

\author{Arsany Basta, Andreas Blenk, \\ Wolfgang Kellerer}
\affiliation{%
	\institution{Technical University of Munich, Germany}
}

\author{Stefan Schmid}
\affiliation{%
	\institution{Aalborg University, Denmark}
}


\begin{abstract}
	Ideally, by enabling multi-tenancy,
	network virtualization allows to improve resource utilization,
	while providing performance isolation: 
	although the underlying resources are shared, the virtual 
	network appears as a 
	dedicated network to the tenant.
	However, providing such an illusion is challenging in practice,
	and over the last
	years, many expedient approaches have been proposed to provide 
	performance isolation in virtual networks, 
	by enforcing bandwidth reservations.
	
	We in this paper study another source
	for overheads and unpredictable 
	performance in virtual networks: the hypervisor.
	The hypervisor is a critical component in multi-tenant 
	environments, 
	but its overhead and influence on performance are hardly understood 
	today.
	In particular, we focus on
	OpenFlow-based virtualized Software Defined Networks (vSDNs).
	Network virtualization is considered a killer application for SDNs:
	a vSDN allows 
	each tenant to flexibly manage its network from 
	a logically centralized perspective,
	via a simple API.
	
	For the purpose of our study, we developed a new benchmarking tool
	for OpenFlow control and data planes,
	enabling high and consistent
	OpenFlow message rates.
	Using our tool, we identify and measure controllable and uncontrollable effects 
	on performance and overhead, including 
	the hypervisor technology, 
	the number of tenants as well as the tenant type, 
	as well as the type of OpenFlow messages. 
\end{abstract}

%
%
%

%
%

%
%


\keywords{SDN; Virtualization; Hypervisor Performance Benchmark}


\maketitle

\section{Introduction}



While virtualization has successfully revamped the server business---virtualization
is arguably the single most important paradigm behind the success of cloud
computing---,
other critical components of distributed systems, such as the network,
have long been treated as second class citizens.
For example, cloud providers hardly offer any guarantees on the
network performance today. 
This is problematic:
to provide a predictable application performance,
 isolation needs to be ensured across \emph{all involved 
components and resources}.
For example,
  cloud-based applications, including batch processing,
streaming, and scale-out databases,
generate a significant amount of network traffic and a considerable
fraction of their runtime is due to network acti\-vi\-ty. 
Indeed,  
several studies have shown the negative
impact network interference can have on the predictability 
of cloud application performance. 

\emph{Network virtualization} promises a more predictable cloud
application performance by providing a unified abstraction 
and performance isolation across
nodes and links, not only within a data center or cloud, 
but for example also in the wide-area network. 
Accordingly, over the last years, several
virtual network abstractions such as \emph{virtual clusters}~\cite{oktopus},
 as well as systems 
such as \emph{Oktopus}~\cite{oktopus}, 
\emph{Proteus}~\cite{proteus}, and \emph{Kraken}~\cite{kraken}, 
have been developed. 

While today, the problem of how to
exploit resource allocation flexibilities and provide isolation 
in the data plane is fairly well-understood~\cite{oktopus,proteus,vnet-sdn-jrex,kraken}, 
we in this paper study 
a less well-understood but critical component
in any network virtualization architecture: \emph{the hypervisor}.
A hypervisor is responsible for the multiplexing, de-multiplexing,
and orchestrating resources across multiple tenants.
For example, the hypervisor performs admission control
which is needed to avoid over-subscription
and provide absolute 
performance guarantees for tenants
sharing a finite infrastructure. 

In particular, and as an important case study, we in this paper focus on 
virtual Software-Defined Networks (vSDNs).
Indeed, network virtualization is considered a killer
application for Software-Defined Networks (SDNs)~\cite{vnet-sdn-jrex,road2sdn}:
By outsourcing and consolidating the control 
over data plane devices (OpenFlow switches) to a logically
centralized software, the so-called controller, 
a vSDN allows 
each tenant to flexibly manage its own virtual network(s),
from 
a logically centralized perspective.
In particular, OpenFlow, the de facto SDN standard,
offers a simple API for installing
packet-forwarding rules, querying traffic statistics, and
learning about topology changes. 

To give an example of the importance of the hypervisor
in vSDNs, we may consider the flow setup process:
an SDN controller needs to react
to a new flow arrival by installing flow rules on the switch
accordingly. In a vSDN, the packet arrival event
and flow rule message are communicated indirectly
between the controller and the OpenFlow switch,
via the hypervisor. Thus,
in a scenario where the hypervisor is overloaded,
undesired latencies may be introduced. 
Another example may arise in the context 
of a load balancing application
which requires link utilization statistics every 10 ms: 
even if a controller can handle high-rate statistics
requests, the hypervisor may only supports one OpenFlow
request per second, which can degrade the application performance.

As we will see in this paper,
the application performance in a multi-tenant SDN
is influenced by several additional aspects. 

\noindent \textbf{Our Contributions}

This paper initiates the study of
sources of overheads and unpredictable
performance in multi-tenant virtual
networks based on SDN. 
We present a novel 
benchmarking tool
for OpenFlow which we developed
for this study, and which is
tailored toward high and consistent
OpenFlow message rates.

We show that 
our benchmark tool can help identify sources
of overheads and bottlenecks as well as properties 
of a vSDN architecture, which we hope in turn can help
developing models and improve the hypervisor
design. 
In particular, 
we identify and 
measure the factors
related to the the 
hypervisor technology and mechanism, 
the number of tenants as well as the tenant type, 
and the type of OpenFlow messages. 


\noindent \textbf{Organization}

The remainder of this paper is organized as follows.
Section~\ref{sec:metho} presents our 
benchmark tool and measurement methodology.
Section~\ref{sec:results} reports on our results.
After reviewing related work in Section~\ref{sec:relwork},
we summarize our contributions and outline future work
in 
Section~\ref{sec:conclusion}.

\section{Measurement Methodologies}\label{sec:metho}

In order to conduct our study of the performance and overheads of
the SDN hypervisor, we first needed to develop a novel and more 
flexible benchmark tool.
This section introduces the \emph{perfbench}
tool, and also presents our methodology
and experimental setup.

\subsection{High Throughput Measurement Tool}

\begin{figure}[!t]
	\centering{
		\includegraphics[width=\columnwidth]{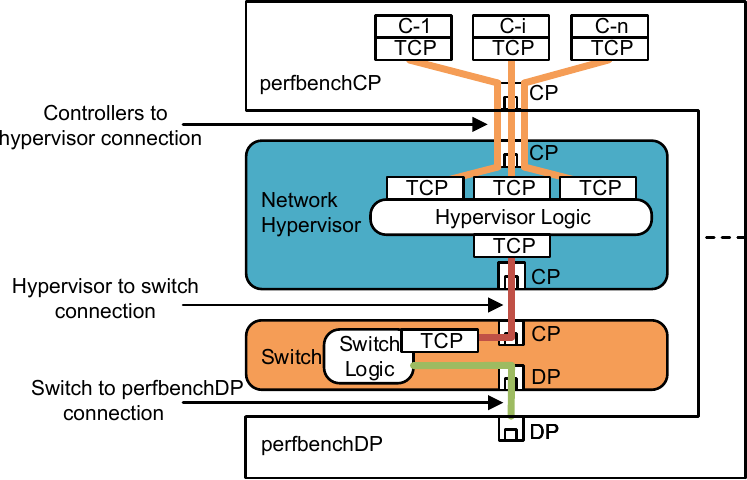} 
	}
	\caption{The \emph{perfbench} Architecture}
	\label{fig:perfbench}
\end{figure}

Due to the incapability to create high and consistent OpenFlow message rates, \emph{perfbench} is designed to emulate high OpenFlow message rates for providing high throughput\--orien\-ted performance benchmarks for SDN networks.
\emph{perfbench} can be used for measurements in multi-tenant as well as non-virtualized SDN networks.
The tools builds on top of libfluid C++ library \cite{libfluid}, which provides the basic implementation and interface of OpenFlow messages. 
It supports OpenFlow versions 1.0 and 1.3.
Figure~\ref{fig:perfbench} provides a simplified view of how \emph{perfbench} is designed and also how it is operated in multi-tenant SDN networks.

\subsubsection{Tool Architecture}
As illustrated, \emph{perfbench} is divided into two parts, a control plane part (\emph{perfbenchCP}) and a data plane part (\emph{perfbenchDP}).
\emph{perfbenchCP} runs the processes for the tenant SDN controllers, i.e., it emulates
the tenant SDN controller.
Each controller is running in its own thread and is connecting via a unique TCP socket to guarantee isolation between the tenant controllers.
For each controller process, a pre-determined constant message rate of OpenFlow messages can be generated: 
given a certain message rate in seconds, 
\emph{perfbench} generates an equal distribution
of requests at millisecond precision. 

\emph{perfbenchDP} runs the data plane part of \emph{perfbench}.
While it can be connected to existing OpenFlow switches directly, e.g., to receive UDP packets sent via \packetout messages, it can also emulate OpenFlow switches.
In case it emulates an OpenFlow switch, it connects directly to the hypervisor's control plane mode.
This operation mode can be useful to just benchmark the real overhead as induced by hypervisors.
In this paper, we are operating \emph{perfbench} only in the first mode, i.e., we are not emulating OpenFlow SDN switches.

\begin{figure}[!t]
	\centering{
		\includegraphics[width=\columnwidth]{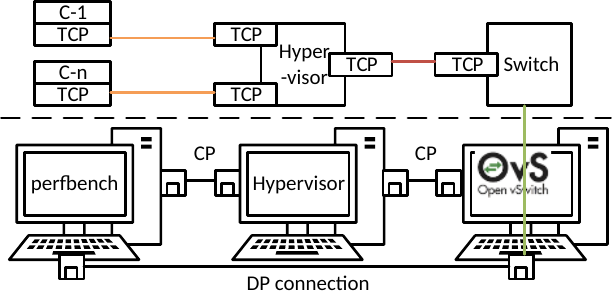} 
	}
	\caption{Experimental Setup}
	\label{fig:setup}
\end{figure}

\begin{table*}
	\centering
	\caption{Measurement setups and Configurations}
	\begin{tabular}{|l|l|l|l|l|l|l|l|} \hline
OF Message Type&Tenants&Switch-only&Message Rate& \tcpnodelay &Hypervisors&No. of runs&Duration \\ \hline
		\packetin & 1 & No &10k,20k,30k,40k & 0 & FV,OVX&10&30~sec \\  \hline
		\portstats & 1 & No & 5k,6k,7k,8k & 0 & FV,OVX&10&30~sec \\  \hline
		\packetin & $2:20$ & Yes & 40K & 0/1 & FV,OVX&10&30~sec \\  \hline
		\packetout & $2:20$ & No & 60K & 0/1 & FV,OVX&10&30~sec \\  \hline
	\end{tabular}
	\label{tab:setups}
\end{table*}

\subsubsection{Supported Message Types \& Latency}
\emph{perfbench} supports the following message types: $\packetin$, \texttt{OFPT\-\_\-PACKET\_\-OUT}, $\echorequest$, $\featuresrequest$, $\portstats$.
For synchronous OpenFlow messages, i.e., where a request expects an answer, e.g., $\portstats$, $\echorequest$, $\featuresrequest$, the latency is measured as the time it takes from sending the request until receiving the reply.

In case of asynchronous OpenFlow messages, namely \packetin and \texttt{PACKET\_OUT}, the latency calculation is slightly different. 
For \texttt{PACKET\_IN}, \emph{perfbenchDP} sends UDP packets for each tenant via its data plane connection.
The latency is then calculated as the time it takes from sending the UDP packet until receiving the \packetin at \emph{perfbenchCP}.

For \texttt{PACKET\_OUT}, \emph{perfbenchCP} triggers the sending of \packetout with artificial data packets. The latency is then calculated for each tenant as the time it takes from sending the \packetout until receiving the artificial data packet at \emph{perfbenchDP}.



Besides, \emph{perfbench} provides the capability to set the \tcpnodelay flag for a specific TCP connection.
Setting \tcpnodelay disables \textsc{Nagle}'s algorithm.
While \textsc{Nagle}'s algorithm has been introduced to improve network performance in general, as we will see, it can
lead to performance costs in case of SDN-based networks.
\textsc{Nagle} is used to aggregate more data, thus produce less packet overhead per TCP packet.
However, this aggregation of packet content might lead to higher latencies per packet.
As SDN application performance can be severely affected by high delays, \textsc{Nagle}'s algorithm hence might lead to performance degradation in SDN networks.
Accordingly, to investigate the impact of \textsc{Nagle}, \emph{perfbench} provides the capability to set \tcpnodelay~.

\subsection{Measurement Setup and Test Cases}
Figure~\ref{fig:setup} shows the measurement setup. 
Three PCs are used to conduct the hypervisor performance benchmarks in this paper.
\emph{perfbench}  (\emph{perfbenchCP} and \emph{perfbenchDP}) runs on the left PC, one hypervisor (FV or OVX) on the middle PC, and an OpenVSwitch (OVS)~\cite{pfaff2009} instance on the right PC.
perfbenchCP is connected to the hypervisor PC.
The hypervisor PC is connected to the OVS PC.
perfbenchDP is connected via a dedicated line to the data plane part of the OVS PC.

For a short representative measurement study, we choose \packetin and \packetout for asynchronous message types, and \portstats for synchronous message types.
\featuresrequest and \texttt{OFPT\-\_ECHO\-\_RE\-QUEST} are neglected as we see them as not critical for the runtime performance of SDN networks.

Table~\ref{tab:setups} provides an overview of all conducted measurements.
For all message types, single tenant (1) as well as multi-tenant (2:20) measurements are conducted for a range of rates, \tcpnodelay settings, and the two hypervisors FlowVisor (FV) and OpenVirteX (OVX).
Every setup is repeated 30 times for a duration of 30 seconds.
As we are interested in the steady-state performance, 
we cut the first and last 5 seconds from the data analysis;
the remaining 20 seconds show a stable pattern.

For the multi-tenancy measurements, the hypervisor instances are configured according to their specificity.
This means, for instance, that for OVX \emph{perfbenchDP} uses artificial unique MAC addresses per tenant as this is a pre-requisite for the operation of OVX.
As FV uses flowspace slicing, such a setting is not necessary.



\section{Measurements and Evaluation}\label{sec:results}

We structure our measurement study into two parts: 
single tenant experiments and multi tenant experiments.
In the first part, we investigate 
how different hypervisor implementations affect the control plane performance,
as well as how the performance depends on the OpenFlow message types.
In the second part, we investigate whether and how 
the control latency depends on the number 
of tenants,
and how the tenants' controller impact the hypervisor performance.
Finally, we take a brief look at fairness aspects.

\subsection{Single Tenant Evaluation}

The hypervisor performance is evaluated in terms of control plane latency, and compared against different control message rates. 
We compare two state-of-the-art 
hypervisor implementations, namely FlowVisor (FV) and OpenVirteX (OVX). 
Moreover, we consider the performance for two 
OpenFlow message types, namely asynchronous and 
synchronous messages (see above). 
We consider asynchronous \packetin 
messages in our 
experiments since their performance is critical for flow setup. 
For synchronous messages, we consider \portstats as an example:
it is used by SDN apps to collect port statistics, 
e.g., for load balancing or congestion-aware routing.

\begin{figure*}[th]
	\centering{
		\subfloat[Hypervisors overhead, Single tenant, \packetin, 40k rate]{
			\includegraphics[width=0.31\textwidth]{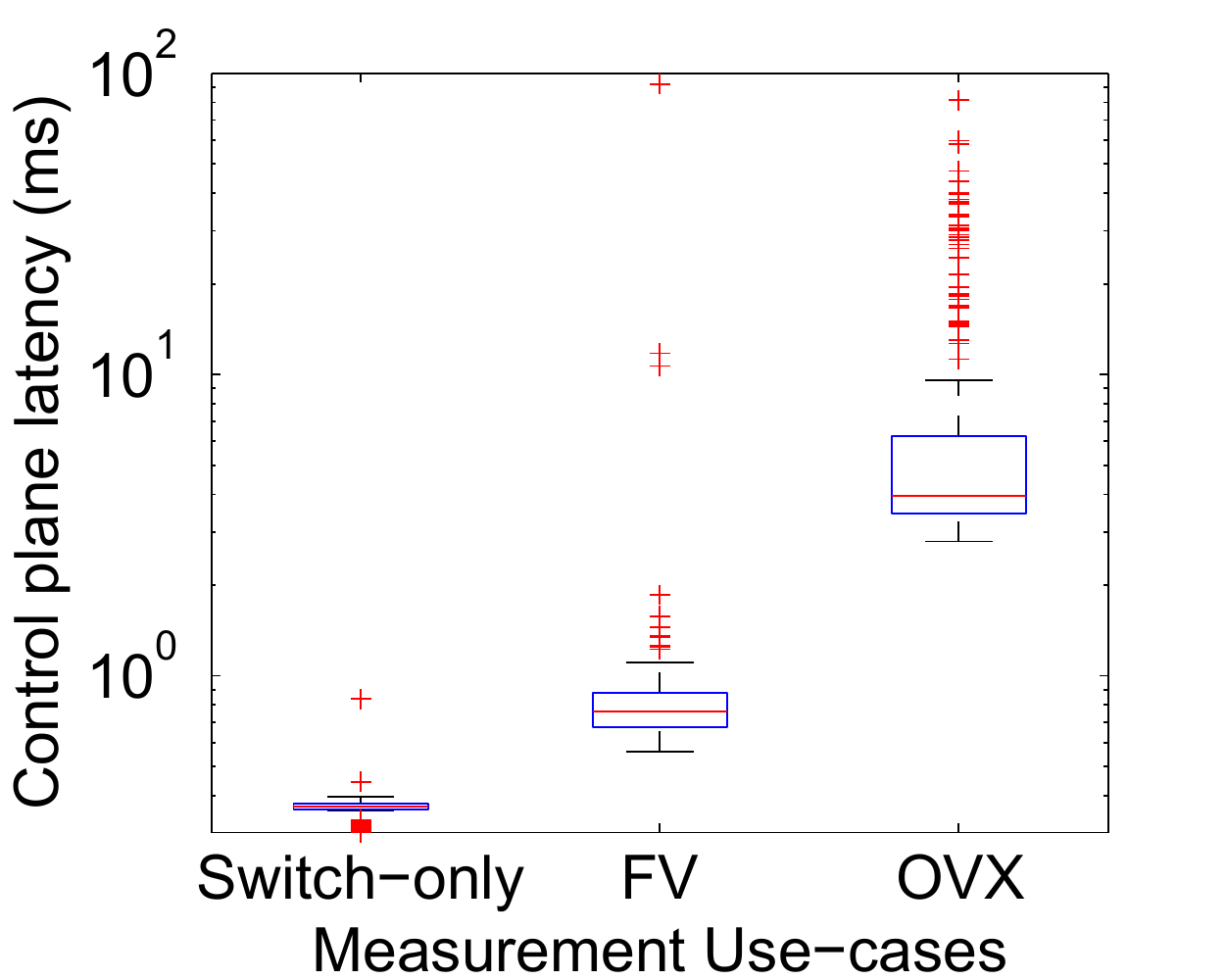} 
		}
		\hfil
		\subfloat[FV vs. OVX, Single tenant, \packetin, 10-40k rate]{
			\includegraphics[width=0.31\textwidth]{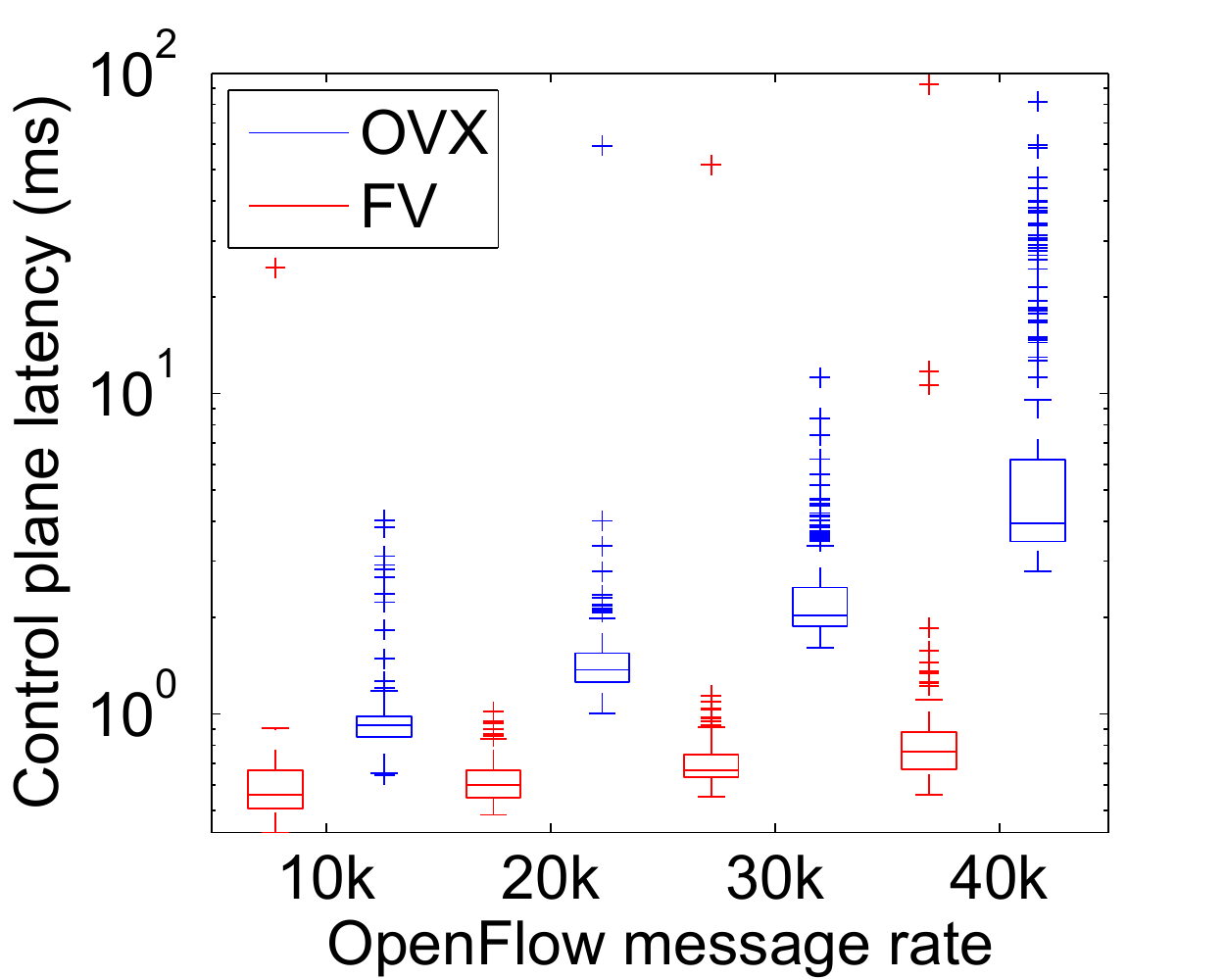} 
		}
		\hfil
			\subfloat[FV vs. OVX, Single tenant, \portstats, 5-8k rate]{
			\includegraphics[width=0.31\textwidth]{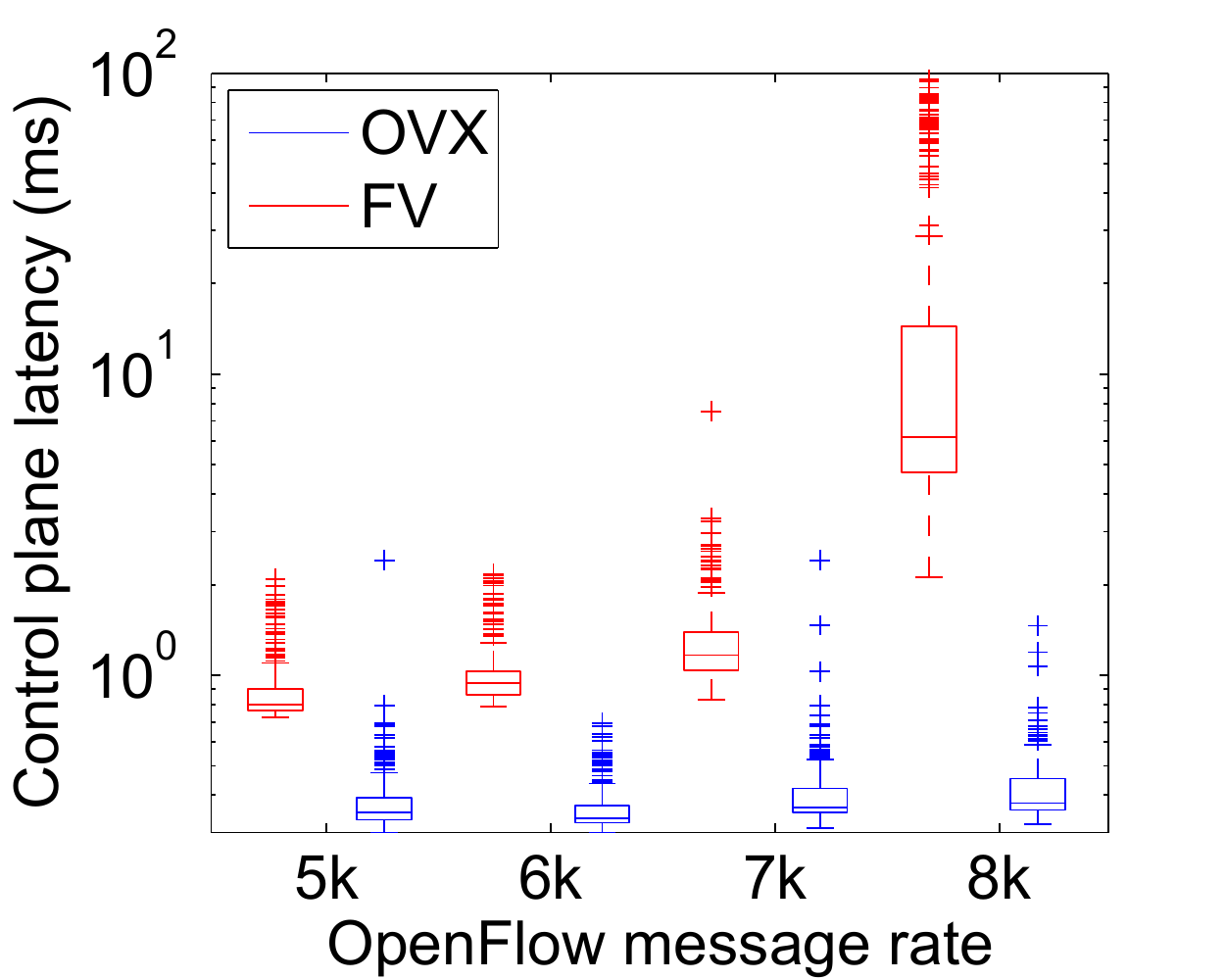} 
		}
	}
	\caption{Single tenant, latency}
	\label{fig:st}
\end{figure*}

Fig.~\ref{fig:st}a shows the control plane performance overhead induced by 
 the indirection via the hypervisor: 
 a ``man-in-the-middle'' between controllers and switches. 
 The evaluation considers a
 setting where \packetin messages are arriving at a rate of 40k per second, 
 which is the maximum rate for this OpenFlow message type 
 that can be generated by our tool on the used computing platform. 
 The control plane performance is considered in terms of the control plane latency, where FV shows an average of 1 ms (millisecond) compared to  
 0.1 ms with the switch-only. OVX adds even more latency overhead with 3 ms compared to an 0.3 with switch-only. The control latency overhead could be observed for both FV and OVX, due to adding extra intermediate network processing.


\textbf{How do different hypervisor implementations affect the control plane performance?} 

In order to evaluate the difference between the hypervisor implementations, we evaluate the observations from the measurements of the \packetin OpenFlow messages, shown in Fig~\ref{fig:st}b. The \packetin message rate is ranging from 10k to 40k messages per second. 

The measurements show that FV features a lower control latency than 
OVX, especially with increasing message rates. OVX shows higher latency and 
more outliers with varying rates due to the control message translation 
process, e.g., an average of 1 ms for 10K up to an average of 3 ms for 40k. This is because OVX includes data plane packet header re-writing from a given virtual IP address specified for each tenant to a physical IP address used in the network. 
Also note the outliers with OVX at 40k, indicating
a possible source of unpredictable performance. In contrast, FV operates in a transparent manner where it does not change the data plane packet headers and it operates with an average of 1 ms control latency for all evaluated rates. The \packetin handling at FlowVisor results in lower control latency and a more robust performance 
under varying control rates.

\textbf{How does the performance change with different OpenFlow message types?} 

 For this evaluation, we also consider a single tenant, however measuring the control latency for \portstats messages. The measurement is carried out
 at message rates between 5k and 8k per second, 
 due to the limits of the \portstats rate the used switch can handle. 
 As shown in Fig.~\ref{fig:st}c, the transparent design shows 
 ineffiency and overhead in terms of control latency for \portstats, e.g., 
 going from an average of 1 ms with 5k up to an average of 7 ms at 8k. 
 Since FV transparentely forwards all message to the switch, the switch can 
 become overloaded, hence, the control latency increases proportionally to the port stats rates. The switch becomes overloaded at a rate of 8k \portstats per second. 
 OVX uses a different implementation for synchronous messages: 
 it does not forward the port stats to the switches, but 
 rather pulls it from the switch 
 given 
 the number per second. OVX replies on behalf of the switch to all other 
 requests, and hence, avoids 
 overloading the switch, resulting 
 in a better control plane latency performance. 
 However, we also note a drop between 5k and 6k for OVX, 
 indicating a source of unpredictability.

%
%


%

\subsection{Multi Tennant Evaluation}

We study how the vSDN performance depends 
on the number of deployed tenants. Recall that
ideally, in a virtual network, 
the performance should not depend on the presence or number of other
tenants.
We also measure the influence of the tenant's controller implementation on the hypervisor performance. For this purpose, we consider two implementations for the tenant's controller considering the packaging of OpenFlow messages to TCP packets. The controller can either aggregate multiple OpenFlow messages in a TCP packet, which we refer in short as (AGG). Alternatively, the controller can exploit the \tcpnodelay setting and send each OpenFlow message once it is generated in a TCP packet, which we refer to by (ND).

\textbf{How does the performance, i.e., control latency, change with increasing number of tennants?} 

For the multi tenant evaluation, we use \packetout OpenFlow messages, 
since they originate 
from the tenant's controller and can be influenced by the controller implementation. We iterate from 2 tenants up to 20 tenants deployed on the hypervisor:
for comparison purposes, we adjust the per-tenant message rate such that the total rate remains constant.
 The \packetout message rate used in this evaluation is 60k messages per second.

\begin{figure}[!t]
	\centering{
		\includegraphics[width=0.38\textwidth]{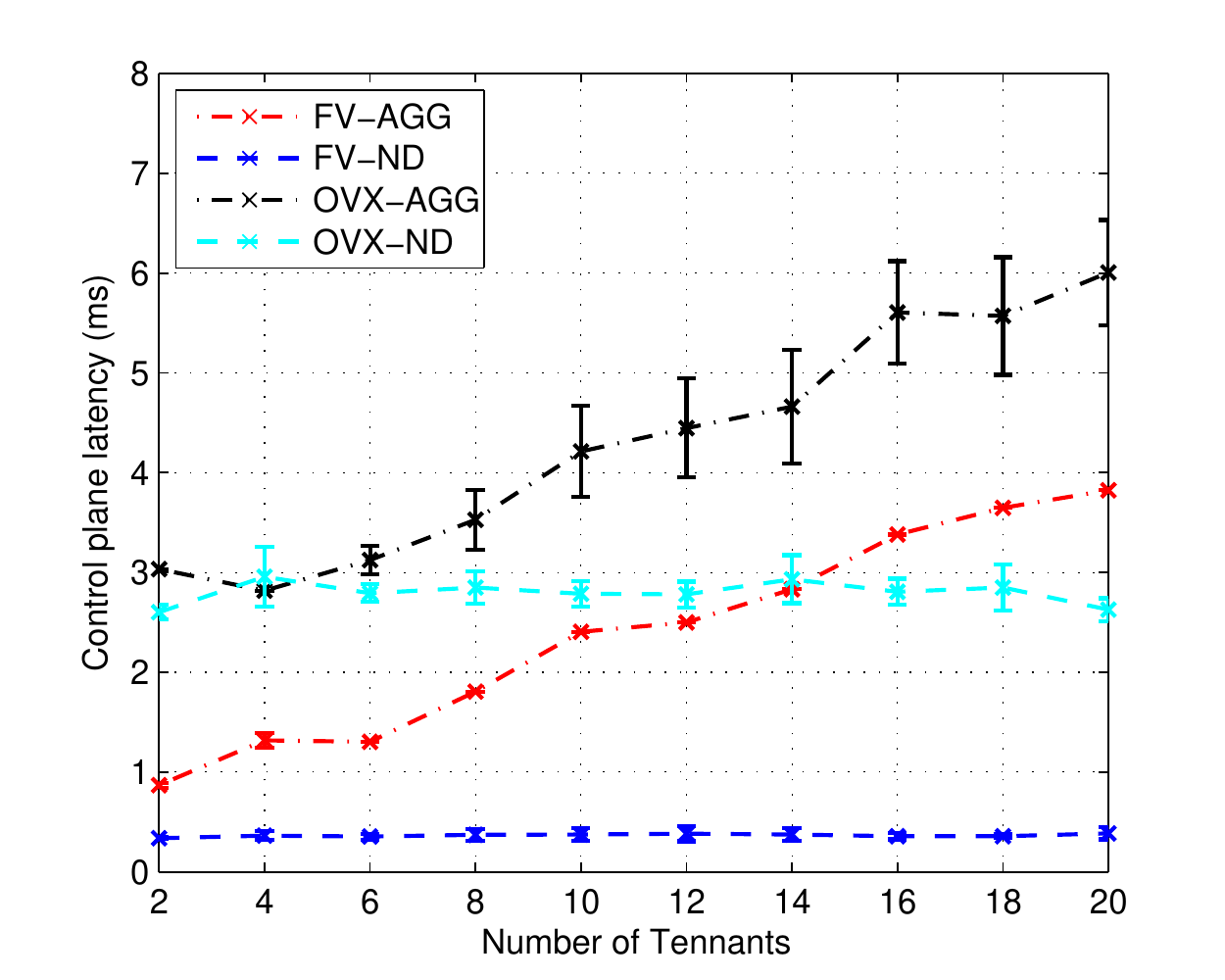} 
	}
	\caption{Multiple tenants, \packetout, 60k rate, latency}
	\label{fig:mt_pcktout_lat}
\end{figure}

The impact of increasing the number of tenants is shown in Fig.~\ref{fig:mt_pcktout_lat}. We discuss first the impact of increasing the tenants on both FV and OVX with the default controller implementation with \tcpnodelay = 0, i.e., aggregation of several OpenFlow messages on the same TCP packet. For both hypervisors, depicted as ``FV-AGG" and ``OVX-AGG", increasing the number of tenants degrades the performance of the control plane and adds more latency overhead. However, this is mainly driven by the setting of the tenant's controller, where the controller adds waiting time till enough OpenFlow messages are there to be sent on a TCP packet. For example with a fixed 60k $\packetout$, at 2 tenants, each tenant generates 30k messages per second, however at 20 tenants, each tenant only generates 6k messages per second. Hence, controller of each tenant at 20 tenants experiences waiting times till enough OpenFlow messages are available to be sent on a TCP packet, i.e., aggregation. This behavior results in control latency of an average 6 ms compared to 3 ms only at 2 tenants, with OVX-AGG for example. 
Another remark is that OVX shows more control plane latency than FV for \packetout messages, similar to the control latency observations for \packetin messages.

\textbf{How does the tenant's controller impact the hypervisor performance?} 

The impact of the tenant's controller implementation is shown in in Fig.~\ref{fig:mt_pcktout_lat}, depicted for both hypervisors as ``FV-ND" and 
`'OVX-ND". Using the \tcpnodelay =1 at the tenant's controller, both hypervisors show a significant improvement compared to the OpenFlow aggregation implementation such that the control latency becomes decoupled from the number of deployed tenants. FV results in
a control latency of, on average, less than 1 ms,
independentl of the number of tenants, while OVX results in 3 ms for all tenants.
The \tcpnodelay setting allows the generated OpenFlow messages to be sent directly, while message aggregation on TCP connection adds to the control latency: 
OpenFlow messages have to wait at the controller after being generated. 
Note that the hypervisor
 cannot control the tenant's controller behavior,
 which introduces a source of unpredictability.

\begin{figure}[!t]
	\centering{
		\includegraphics[width=0.38\textwidth]{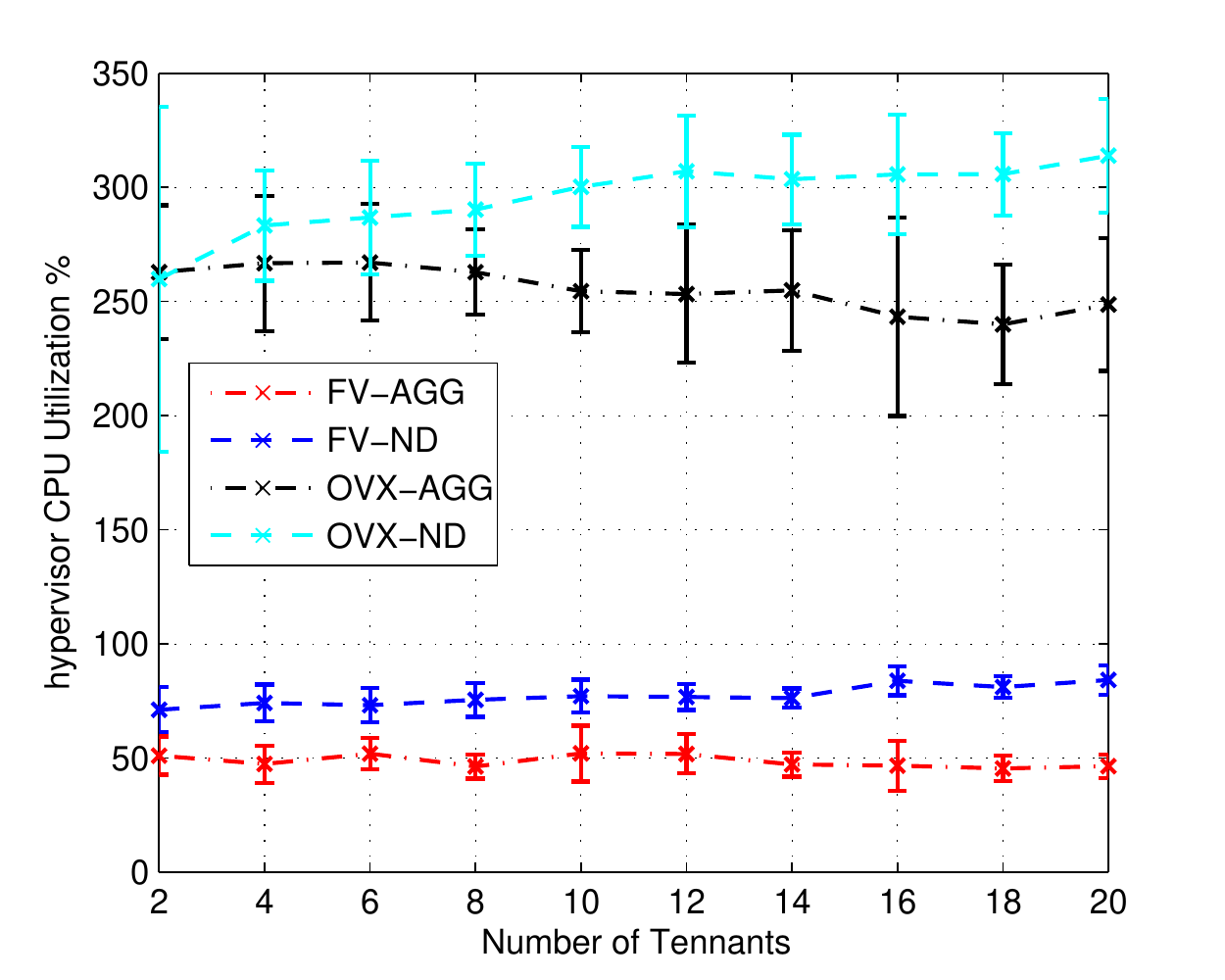} 
	}
	\caption{Multiple tenants, \packetout, 60k rate, CPU}
	\label{fig:mt_pcktout_cpu}
\end{figure}

The workload of the hypervisors, in terms of CPU utilization, for both FV and OVX is shown in~Fig.~\ref{fig:mt_pcktout_cpu}. Note that OVX is multi-threaded, hence can utilize more than 1 CPU core, compared to FV which is only single threaded. The first insight is that FV requires much less CPU to process the same OpenFlow packet type and rate, e.g., 50\% of 1 CPU core with aggregation, in this 
setting at a \packetout rate of 60k per second. Considering the difference in the CPU utilization, comparing the \tcpnodelay setting, the CPU utilization is higher compared to aggregation, for both FV and OVX. For example, OVX utilizes 50\% more CPU at 20 tenants with \tcpnodelay = 1. It is intuitive to see that in case \tcpnodelay flag is enabled, more TCP packets are generated by the tenant and have to be processed by the hypervisor which increases the hypervisor's CPU load.

\textbf{How is the observed control latency distributed among the multi tennants, i.e., fairness?} 

\begin{figure*}[th]
	\centering{
		\subfloat[FV, \tcpnodelay = 0]{
			\includegraphics[trim={5pt 0pt 20pt 15pt},clip,width=0.22\textwidth]{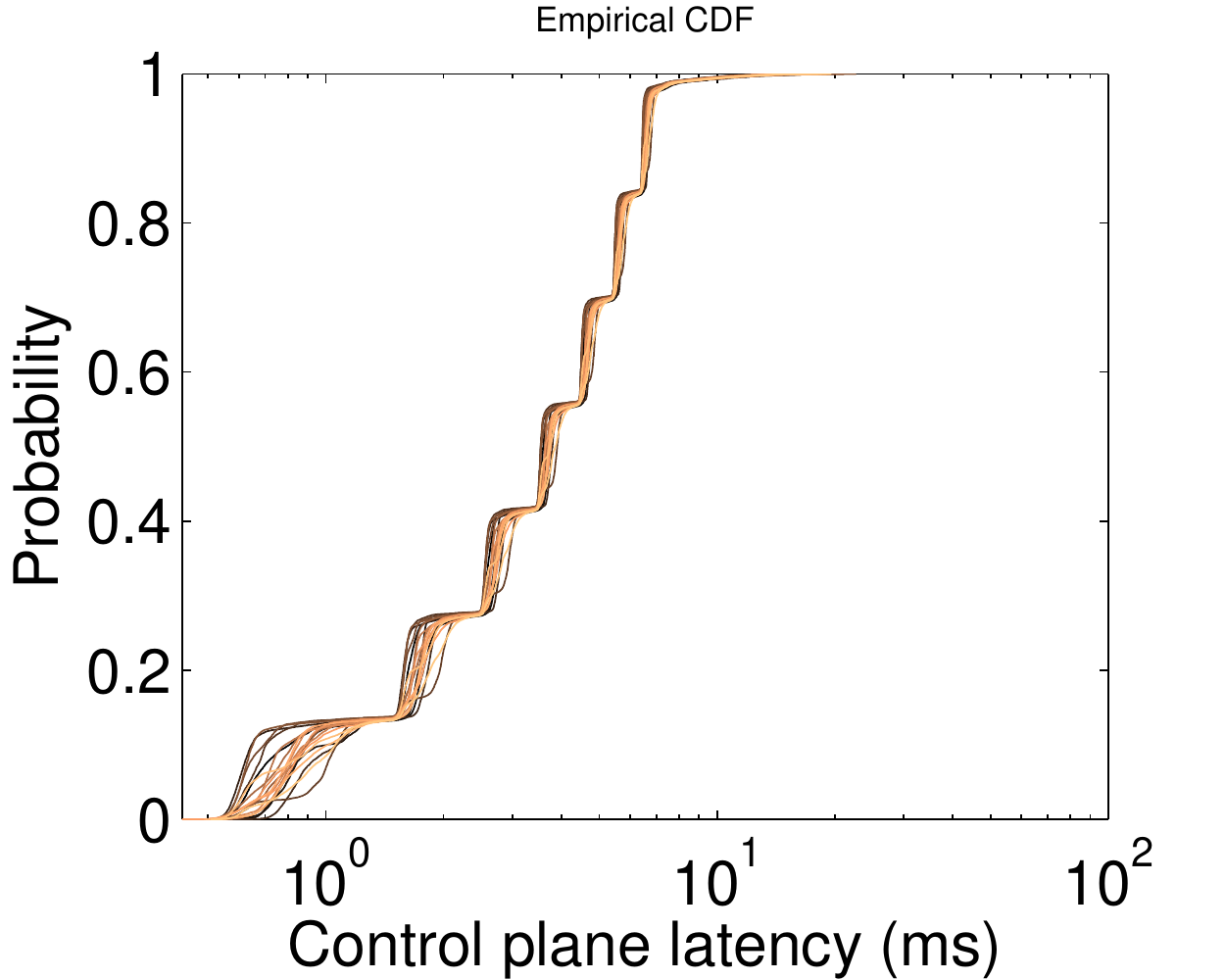} 
		}
		\hfil
		\subfloat[FV, \tcpnodelay = 1]{
			\includegraphics[trim={5pt 0pt 20pt 15pt},clip,width=0.22\textwidth]{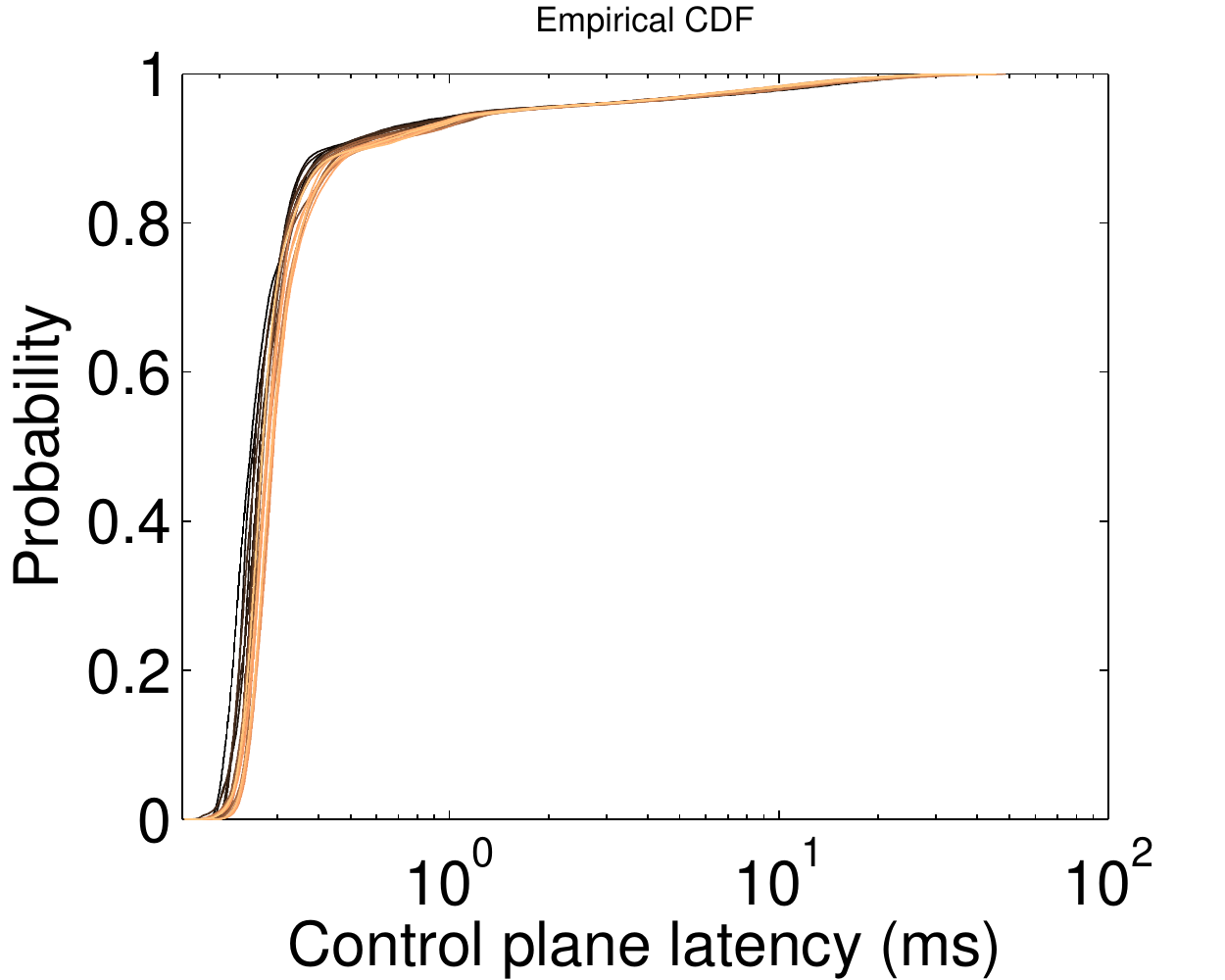} 
		}
		\hfil
			\subfloat[OVX, \tcpnodelay = 0]{
			\includegraphics[trim={5pt 0pt 20pt 15pt},clip,width=0.22\textwidth]{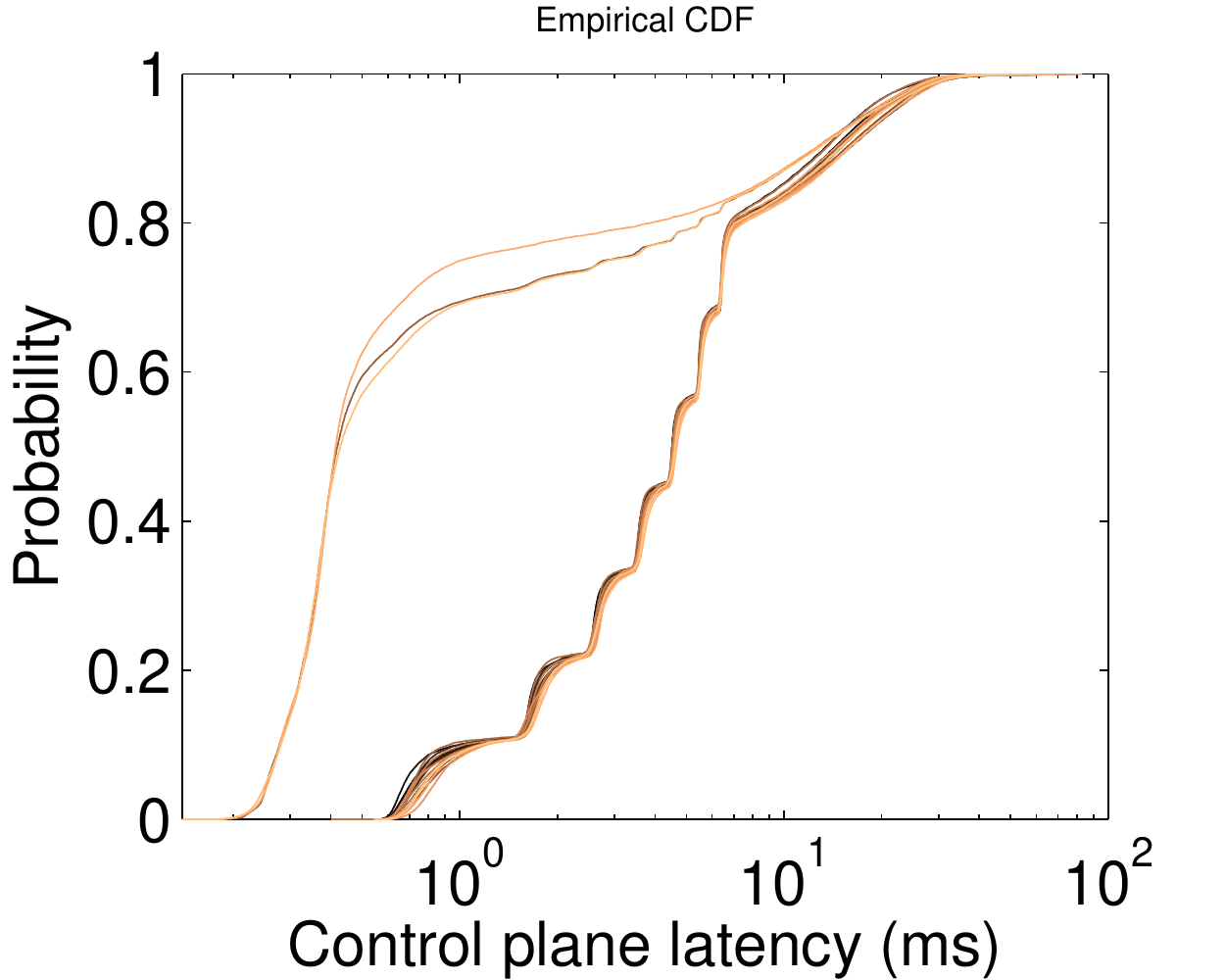} 
		}
		\hfil
			\subfloat[OVX, \tcpnodelay = 1]{
			\includegraphics[trim={5pt 0pt 20pt 15pt},clip,width=0.22\textwidth]{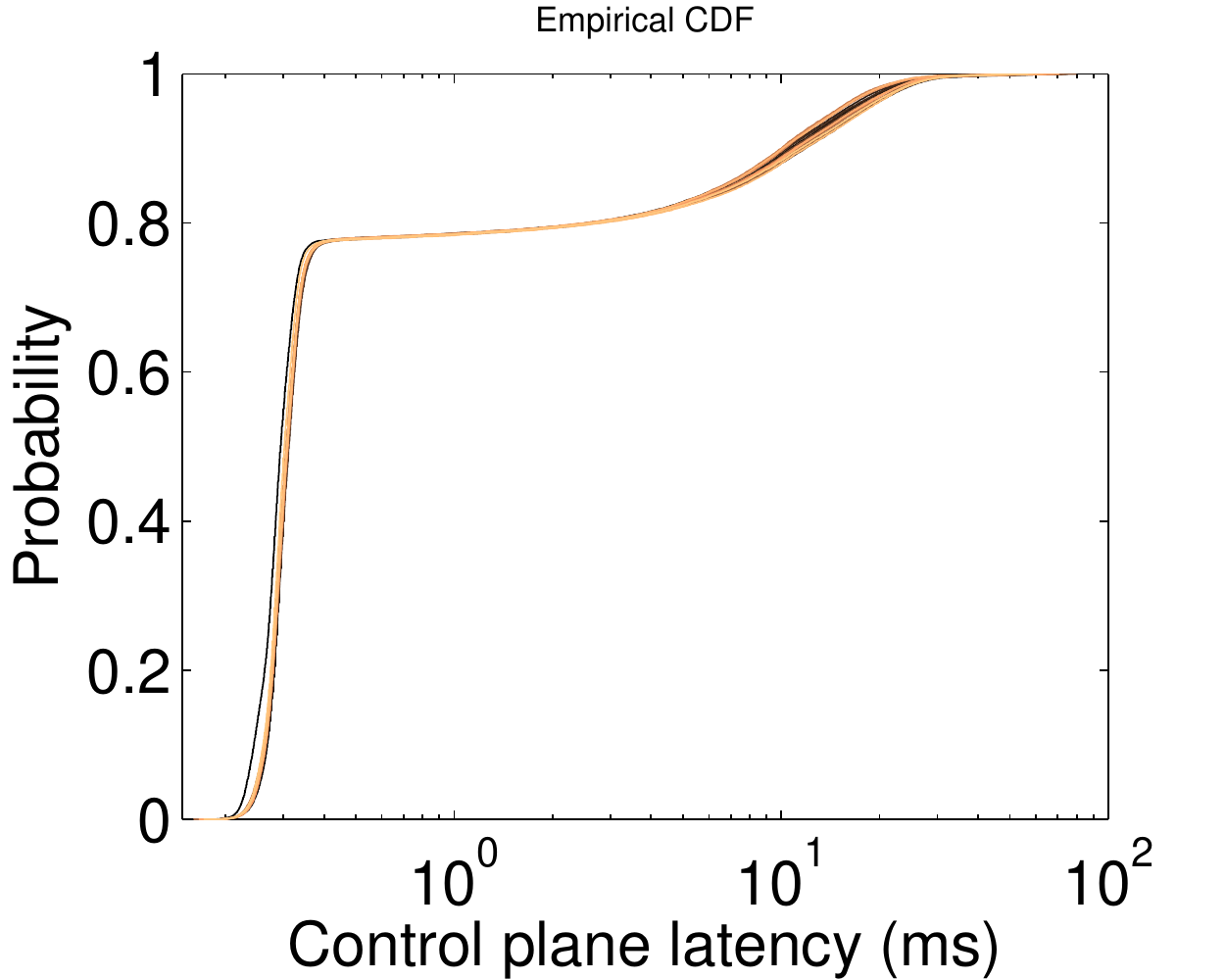} 
		}
	}
	\caption{20 Tenants, \packetout, 60k rate, Latency distribution for each tenant}
	\label{fig:fairness}
\end{figure*}

In order to investigate the perforamnce impact on individual tenants, we measure the latency per tenant for the setup with \packetout, with 60k rate and for 20 tenants, i.e., max setup/settings. The control plane latency distribution over a single run is shown in fig~\ref{fig:fairness} for both hypervisors and \tcpnodelay = 0 and = 1. 

In general, we could observe fair latency distribution among all 20 tenants, except for the the case with OVX and aggregation, in Fig.~\ref{fig:fairness}c. There are 3 out of 20 tenants which experience a control latency with an average of 0.5 ms, while all other tenants experience an average control latency of 6 ms.
This defines the control latency guarantees that can be provided by the hypervisor, 
which requires considering the worst not the best latency performance.
This can result in unpredictabiliy and unfairness.

\section{Related Work}\label{sec:relwork}

There exists a large body of literature on overheads and sources of unpredictable performance in cloud applications. For example,  
several studies have reported on the significant variance of
the bandwidth available to tenants in the absence of
network virtualization: the bandwidth may very by a factor of five or
more~\cite{amazonbw}, even within the same day. 
Given the time spent
in network activity by these applications, this variability has a
non-negligible impact on the application performance, which makes it
impossible for tenants to accurately estimate the execution time in
advance.
Accordingly, over the last years, 
many network virtualization architectures and prototypes have been
proposed, leveraging admission control and bandwidth reservations
and 
enabling tenants to specify absolute
guarantees~\cite{oktopus,ccr-vc,secondnet,faircloud,gatekeeper,blender,proteus}.

There already exists a large body of literature
on hypervisors as well. 
We in this paper are particularly interested in
hypervisors for SDN, and we refer the reader to Blenk et al.~\cite{netvirthyper}
for a good survey.
Existing SDN hypervisors 
can be classified into two categories: centralized (e.g.,~\cite{flowvisor})
and distributed (e.g.,~\cite{vnet-sdn-jrex}). 
FlowVisor~\cite{flowvisor} is one of
the most well-known hypervisors today. FlowVisor
assigns different tenants to different sub-spaces of
the header field space (so-called flow spaces),
and provides isolation (both in terms of address space
as well as in terms of resources) in the data plane.
FlowVisor
has already been extended
in several directions, e.g., with an intermediate control
plane slicing layer that contains a \emph{Flowspace Slicing Policy
(FSP)}~\cite{fsp} engine, or
with improved abstraction mechanisms~\cite{vertigo,advisor}.
\emph{Enhanced FlowVisor}~\cite{eflowvisor},
based on NOX,
adds bandwidth reservations (using VLAN PCP)
and admission control. 
\emph{Slices Isolator}~\cite{siso} is positioned between
the physical SDN network and the virtual SDN controllers,
and allows to adapt the isolation demands of the virtual network
users. 
\emph{FlowN}~\cite{vnet-sdn-jrex} is the first 
distributed hypervisor for virtualizing
SDN networks. It is based on container virtualization and 
does not provide tenants 
with their 
own virtual SDN controller. 
Instead of only slicing the physical network, FlowN completely
abstracts the physical network and provides virtual
network topologies to the tenants. 
\emph{OpenVirteX}~\cite{openvirtex} 
provides address and topology
virtualization, by operating as an intermediate layer between
the virtual SDNs and controllers.

Interestingly, although a hypervisor lies at the heart
of any multi-tenant and network virtualized system, 
the hypervisor and especially its performance and possible
overheads have received little attention:
a gap which we aim to fill with our paper.
Indeed, the survey by Blenk et al.~\cite{netvirthyper}
states a comprehensive
performance evaluation framework 
as a main
open problem for future research. 

Finally, there also exists a comprehensive list
of literature on OpenFlow performance and measurements. 
For example, Hendriks et al.~\cite{hendriksassessing} 
consider the suitability of OpenFlow as a traffic measurement
tool (see~\cite{measure-sdn} for a
survey on the topic), and show that the quality of actual measured data
can be questionable: The authors demonstrate that inconsistencies and
measurement artifacts can be found due to particularities of
different OpenFlow implementations, making it impractical to
deploy an OpenFlow measurement-based approach in a network
consisting of devices from multiple vendors. In addition, they
show that the accuracy of measured packet and byte counts
and duration for flows vary among the tested devices, and in
some cases counters are not even implemented for the sake of
forwarding performance.
Also other authors observed inconsistencies between
bandwidth measurements results and and
a packet-based ground truths~\cite{OpenNetMon}. 
OpenFlow monitoring systems are implemented
similarly to NetFlow, 
and accordingly, problems regarding 
insufficient timestamp resolution~\cite{oneway,peeling}, 
and device artifacts~\cite{uncovering} 
also apply. 
Finally,
Ku{\'z}niar et al.~\cite{kuzniar2015you} 
report on the performance characteristics of flow table updates in 
different
hardware OpenFlow switches, and highlight differences
between the OpenFlow specification and its implementations,
which may threaten correctness
or even network security.

\section{Conclusions}\label{sec:conclusion}

We in this paper initiated the empirical study
of 
performance costs related to the
hypervisor in an 
SDN-based virtual network.
We argued that the hypervisor is a critical
but not well-understood
component in any network virtualization
environment supporting multi-tenancy
in general,
and in SDNs in particular:
as requests
from the controller and replies as well as notifications from the 
OpenFlow devices
have to pass through the hypervisor, the 
tenant's application performance is highly 
influenced by the performance
of the hypervisor and its workload. 



\noindent \textbf{ACKNOWLEDGMENT.}
Research supported by the European Research Council (ERC) 
project FlexNets and Aalborg University's talent management programme
(project PreLytics). 

{\small
%
\bibliographystyle{abbrv}
\bibliography{sigproc}  

\begin{thebibliography}{10}

\bibitem{libfluid}
libfluid.
\newblock {http://opennetworkingfoundation.github.io/libfluid/}.

\bibitem{openvirtex}
A.~Al-Shabibi, M.~De~Leenheer, M.~Gerola, A.~Koshibe, G.~Parulkar,
  E.~Salvadori, and B.~Snow.
\newblock Openvirtex: Make your virtual sdns programmable.
\newblock In {\em Proceedings of the workshop on Hot topics in Software Defined
  Networking (HotSDN)}, pages 25--30. ACM, 2014.

\bibitem{fsp}
C.~Argyropoulos, S.~Mastorakis, K.~Giotis, G.~Androulidakis, D.~Kalogeras, and
  V.~Maglaris.
\newblock Control-plane slicing methods in multi-tenant software defined
  networks.
\newblock In {\em Proc. IFIP/IEEE IM}, pages 612--618. IEEE, 2015.

\bibitem{oktopus}
H.~Ballani, P.~Costa, T.~Karagiannis, and A.~Rowstron.
\newblock Towards predictable datacenter networks.
\newblock In {\em ACM SIGCOMM Computer Communication Review}, volume~41, pages
  242--253. ACM, 2011.

\bibitem{netvirthyper}
A.~Blenk, A.~Basta, M.~Reisslein, and W.~Kellerer.
\newblock Survey on network virtualization hypervisors for software defined
  networking.
\newblock {\em {IEEE} Communications Surveys and Tutorials}, 18(1):655--685,
  2016.

\bibitem{uncovering}
{\'I}.~Cunha, F.~Silveira, R.~Oliveira, R.~Teixeira, and C.~Diot.
\newblock Uncovering artifacts of flow measurement tools.
\newblock In {\em International Conference on Passive and Active Network
  Measurement (PAM)}, pages 187--196. 2009.

\bibitem{vertigo}
R.~Doriguzzi~Corin, M.~Gerola, R.~Riggio, F.~De~Pellegrini, and E.~Salvadori.
\newblock Vertigo: Network virtualization and beyond.
\newblock In {\em Proc. IEEE EWSDN}, 2012.

\bibitem{vnet-sdn-jrex}
D.~Drutskoy, E.~Keller, and J.~Rexford.
\newblock Scalable network virtualization in software-defined networks.
\newblock {\em IEEE Internet Computing}, 17(2):20--27, 2013.

\bibitem{siso}
M.~El-Azzab, I.~L. Bedhiaf, Y.~Lemieux, and O.~Cherkaoui.
\newblock Slices isolator for a virtualized openflow node.
\newblock In {\em Proc. NCCA}, 2011.

\bibitem{road2sdn}
N.~Feamster, J.~Rexford, and E.~Zegura.
\newblock The road to sdn.
\newblock {\em Queue}, 11(12):20, 2013.

\bibitem{kraken}
C.~Fuerst, S.~Schmid, L.~Suresh, and P.~Costa.
\newblock Kraken: Online and elastic resource reservations for multi-tenant
  datacenters.
\newblock In {\em IEEE Conference on Computer Communications INFOCOM}, pages
  1--9. IEEE, 2016.

\bibitem{secondnet}
C.~Guo, G.~Lu, H.~J. Wang, S.~Yang, C.~Kong, P.~Sun, W.~Wu, and Y.~Zhang.
\newblock {SecondNet}: {A} data center network virtualization architecture with
  bandwidth guarantees.
\newblock In {\em Proc. ACM CoNEXT}, 2010.

\bibitem{hendriksassessing}
L.~Hendriks, R.~d.~O. Schmidt, R.~Sadre, J.~A. Bezerra, and A.~Pras.
\newblock Assessing the quality of flow measurements from openflow devices.
\newblock 2016.

\bibitem{oneway}
J.~K{\"o}gel.
\newblock One-way delay measurement based on flow data: Quantification and
  compensation of errors by exporter profiling.
\newblock In {\em International Conference on Information Networking (ICOIN)},
  pages 25--30. IEEE, 2011.

\bibitem{kuzniar2015you}
M.~Ku{\'z}niar, P.~Pere{\v{s}}{\'\i}ni, and D.~Kosti{\'c}.
\newblock What you need to know about sdn flow tables.
\newblock In {\em International Conference on Passive and Active Network
  Measurement (PAM)}, pages 347--359. Springer, 2015.

\bibitem{faircloud}
{L. Popa et al.}
\newblock {FairCloud: Sharing the Network in Cloud Computing}.
\newblock In {\em Proc. ACM SIGCOMM}, 2012.

\bibitem{eflowvisor}
S.~Min, S.~Kim, J.~Lee, B.~Kim, W.~Hong, and J.~Kong.
\newblock Implementation of an openflow network virtualization for
  multi-controller environment.
\newblock In {\em IEEE International Conference on Advanced Communication
  Technology (ICACT)}, pages 589--592. IEEE, 2012.

\bibitem{pfaff2009}
B.~Pfaff, J.~Pettit, T.~Koponen, K.~Amidon, M.~Casado, and S.~Shenker.
\newblock Extending networking into the virtualization layer.
\newblock In {\em ACM Workshop on Hot Topics in Networks (HotNets-VIII)}, 2009.

\bibitem{gatekeeper}
H.~Rodrigues, J.~R. Santos, Y.~Turner, P.~Soares, and D.~Guedes.
\newblock Gatekeeper: Supporting bandwidth guarantees for multi-tenant
  datacenter networks.
\newblock In {\em Proc. WIOV}, 2011.

\bibitem{ccr-vc}
M.~Rost, C.~Fuerst, and S.~Schmid.
\newblock Beyond the stars: Revisiting virtual cluster embeddings.
\newblock In {\em ACM SIGCOMM CCR}, 2015.

\bibitem{advisor}
E.~Salvadori, R.~Doriguzzi~Corin, A.~Broglio, and M.~Gerola.
\newblock Generalizing virtual network topologies in openflow-based networks.
\newblock In {\em Proc. IEEE GLOBECOM}, 2011.

\bibitem{flowvisor}
R.~Sherwood, G.~Gibb, K.-K. Yap, G.~Appenzeller, M.~Casado, N.~McKeown, and
  G.~Parulkar.
\newblock Flowvisor: A network virtualization layer.
\newblock {\em OpenFlow Switch Consortium, Tech. Rep}, pages 1--13, 2009.

\bibitem{peeling}
B.~Trammell, B.~Tellenbach, D.~Schatzmann, and M.~Burkhart.
\newblock Peeling away timing error in netflow data.
\newblock In {\em International Conference on Passive and Active Network
  Measurement (PAM)}, pages 194--203. Springer, 2011.

\bibitem{OpenNetMon}
N.~L. Van~Adrichem, C.~Doerr, and F.~A. Kuipers.
\newblock Opennetmon: Network monitoring in openflow software-defined networks.
\newblock In {\em IEEE Network Operations and Management Symposium (NOMS)},
  pages 1--8. IEEE, 2014.

\bibitem{blender}
K.~C. Webb, A.~Roy, K.~Yocum, and A.~C. Snoeren.
\newblock {Blender: Upgrading Tenant-based Data Center Networking}.
\newblock In {\em Proc. ANCS}, 2014.

\bibitem{proteus}
D.~Xie, N.~Ding, Y.~C. Hu, and R.~Kompella.
\newblock The only constant is change: incorporating time-varying network
  reservations in data centers.
\newblock volume~42, pages 199--210. ACM, 2012.

\bibitem{measure-sdn}
A.~Yassine, H.~Rahimi, and S.~Shirmohammadi.
\newblock Software defined network traffic measurement: Current trends and
  challenges.
\newblock {\em IEEE Instrumentation \& Measurement Magazine}, 18(2):42--50,
  2015.

\bibitem{amazonbw}
{Measuring EC2 system performance}.
\newblock http://goo.gl/V5zhEd.

\end{thebibliography}
}
%
%
\end{document}